\documentclass[letterpaper, 10 pt, conference]{ieeeconf}  

\IEEEoverridecommandlockouts                              

\usepackage{graphicx}
\usepackage{amsmath}
\title{\LARGE \bf
Modeling pressure distribution and heat in the body tissue and extract the relationship between them in order to improve treatment planning in HIFU  
}

\author{Saeed Reza Hajian $^{1}$, Ali Abbaspour Tehrani Fard $^{2}$, Majid Pouladian $^{3,*}$ \\ and Gholam Reza Hemmasi $^{4}$
\thanks{$^{1}$ Medical Radiation Eng. Department, Engineering Faculty, Science and Research Branch, Islamic Azad University, Tehran, Iran}
\thanks{$^{2}$ Assistant Professor, Department of Electrical Engineering, Sharif University of Technology, Tehran, Iran}
\thanks{$^{3}$ Chief of Biomedical Eng. Faculty, Science and Research Branch, Islamic Azad, University, Tehran, Iran
	}
\thanks{$^{4}$ Assistant professor, Research Centre of Gastroenterology \& Hepatology of Firouzgar Hospital, Iran University of Medical Science, Tehran, Iran}
\thanks{$^{*}$ Corresponding Author: Majid Pouladian, Head of Biomedical  Engineering Faculty, Science and Research Branch, Islamic Azad University,Tehran (Iran)}
\thanks{\textbf{Keywords: \textit{HIFU, simulator HIFU, pressure and heat, mechanical model, electrical model, viscoelasticity, voxel, ultrasound elastography}} } 
}

\begin{document}

\maketitle
\thispagestyle{empty}
\pagestyle{empty}

\begin{abstract}

In high intensity focused ultrasound (HIFU) systems using non-ionizing methods in cancer treatment, if the device is applied to the body externally, the HIFU beam can damage nearby healthy tissues and burn skin due to lack of knowledge about the viscoelastic properties of patient tissue and failure to consider the physical properties of tissue in treatment planning. Addressing this problem by using various methods, such as MRI or ultrasound, elastography can effectively measure visco-elastic properties of tissue and fits within the pattern of stimulation and total treatment planning. In this paper, in a linear path of HIFU propagation, and by considering the smallest part of the path, including voxel with three mechanical elements of mass, spring and damper, which represents the properties of viscoelasticity of tissue, by creating waves of HIFU in the wire environment of MATLAB mechanics and stimulating these elements, pressure and heat transfer due to stimulation in the hypothetical voxel was obtained. Through the repeatability of these three-dimensional elements, tissue is created. The measurement was performed on three layers. The values of these elements for liver tissue and kidney of sheep in a practical example and outside the body are measured, and pressure and heat for three layers of liver and kidney tissue of an organism were obtained by applying ultrasound signals with a designed model. This action is repeated in three different directions, and the results are then compared with simulation software for ultrasound, as a reference to U.S. Food and Drug Administration (FDA) measures for HIFU, as well as comparisons of results with an operational method for an HIFU cell. The temperature of modeling on the liver for the practical mode in the first and third layers is 17, 16, and 24 percent, and for the software simulator of the HIFU, the measures are 12.9, 17.9, and 15 percent relative absolute changes. The results for kidney tissue for the layers mentioned is 6, 5.7, and 14.5 percent for the simulator of the HIFU, and 4, 5, and 14 percent compared to the practical mode, demonstrating relative absolute changes. The percentage of absolute changes in pressure for liver and kidney tissue in conducted simulation for the simulator of HIFU also gained 9 percent. It was also observed that treatment planning using the properties of visco-elasticity are especially effective based upon experiments conducted as part of this study.

\end{abstract}

\section{Introduction}
\begin{figure}[h!]
	\includegraphics[width=\linewidth]{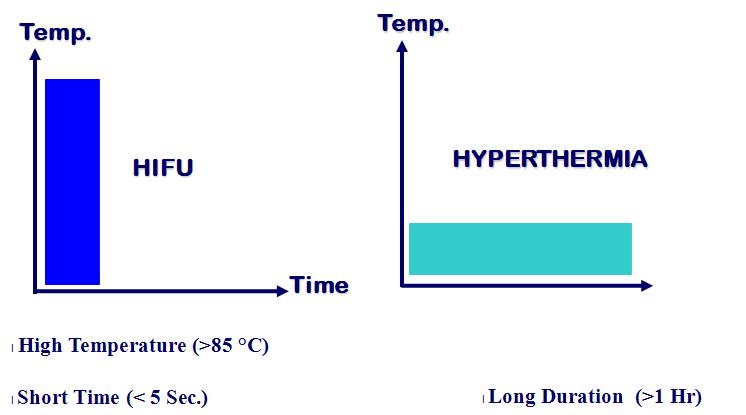}
	\caption{Difference in heat in normal hyperthermia and HIFU}
	\label{fig1}
\end{figure}

Cancer treatment methods can be categorized in two ways: noninvasive and invasive. HIFU is a method of treatment that is primarily used for the removal of solid tumors by creating heat. This method is completely non-invasive. As a radiosurgery procedure it does not damage tissue, and treatment can be performed in one or more sessions. This treatment is done in two ways. In the first method, a probe is inserted through the transrectal area into the desired tissue. The second method is extracorporeal. HIFU waves are radiated from outside into the tissue of the tumor. This device follows standards approved by FDA and CE which stands for European Conformity. Today in different countries, HIFU centers are being established with this system, which is widely used in the treatment of prostate, brain and breast cancer \cite{kremkau1979cancer}. 
As seen in Figure \ref{fig1}, the use of this way compared to general hyperthermia has the advantages that including them can refer to less painful time and high heat in the tumor necrosis \cite{randal2002high}.
\\
Today, this method is used as an extracorporeal treatment for cancers of the liver, pancreas, prostate and bone. As seen in Figure 2, the HIFU method, when using the extracorporeal method, can be effective in the treatment of cancers of the soft tissue by focusing on sound waves.
\begin{figure}[h!]
	\includegraphics[width=\linewidth]{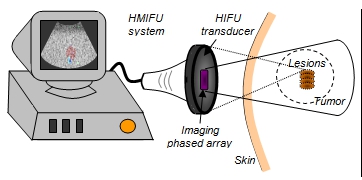}
	\caption{Block diagram using HIFU in destroying the interior of the tumor}
	\label{fig2}
\end{figure}
HIFU (High Intensity Focused ultrasound)  has been studied by numerous researchers as a non-invasive treatment for many clinical problems that require surgery. HIFU is a modern medical method that is bloodless, and portable, and it requires no general anesthesia. This method has limitations, such as the potential to enlarge the size of the lesions created, the need for a cooling system, the inability to treatment by air or bone, problems associated with monitoring of treatment during the healing process, and the inability to plan before treatment \cite{jenne1997ct} \cite{amin2005hifu} \cite{suri2001two}.
Including HIFU applications can be referred for the eradication of cancerous tumors in the prostate, uterus, breast, liver and intestines
\cite{cheng2015contrast} \cite{zhou2014high}. Eradication of a brain tumor is done through the skull as an aid to surgery or minimally invasive surgery usually by means of HIFU applications \cite{sarraf2016deepad} \cite{sarraf2016classification} \cite{sarraf2014brain} 
Recently, many researchers have shown interesting for new applications of HIFU which include cardiac disease such as coronary artery stenosis, cardiomyopathy, hypertrophy and other disorders of the atrial septum, open blood-brain barrier, and fetal tissue \cite{shui2015high}
Clement McDonald and his colleagues, by analyzing thermal imaging data based on two magnetic images, predicted the effect of HIFU in the treatment of a number of fibroids in different areas that are visible with imaging after treatment \cite{imankulov2015feasibility} \cite{saverino2016associative} \cite{sarraf2016robust}.

\section{Heating Biological Model}
\begin{figure}[h!]
	\includegraphics[width=\linewidth]{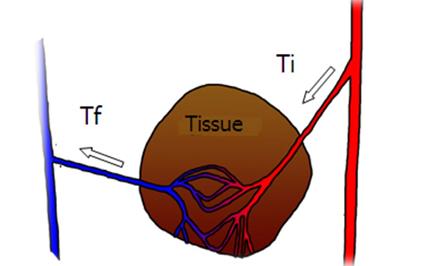}
	\caption{Thermal energy input and output balance of tissue.}
	\label{fig3}
\end{figure}

\begin{figure}[h!]
	\includegraphics[width=\linewidth]{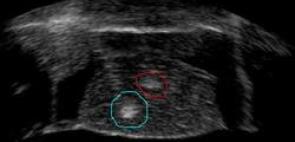}
	\caption{Treatment done in vivo tissue with HIFU and treatment outcomes.}
	\label{fig4}
\end{figure}

\begin{figure}[h!]
	\includegraphics[width=\linewidth]{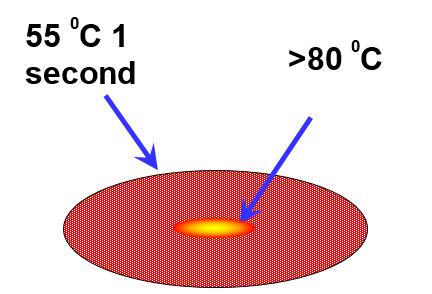}
	\caption{Treatment done in vivo tissue with HIFU and treatment outcomes.}
	\label{fig5}
\end{figure}

\begin{figure}[h!]
	\includegraphics[width=\linewidth]{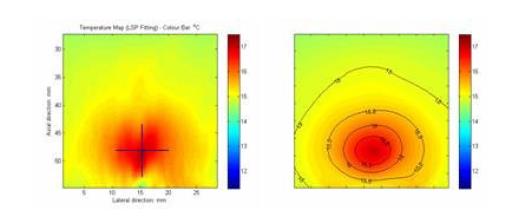}
	\caption{Treatment done in vivo tissue with HIFU and treatment outcomes.}
	\label{fig6}
\end{figure}

A heating biological model is defined by equation \ref{eq1}, through which tissue can be calculated during treatment. In this model, the geometry of tissue and the heat created by absorption of HIFU should be fully studied; Full details of the heat capacity of the tissue should be revealed.
\begin{equation}
\label{eq1}
Q=\frac{-K.A. (\Delta T). (\Delta t)}{(\Delta L)}
\end{equation}
\\
The flow of heat passing through tissue obtained using the above equation is as follows:
\begin{equation}
\label{eq2}
f=\frac{-K.\Delta T}{\Delta L}
\end{equation}
In these equations, $f$ is heat flow, $k$ is weakened heat conduction, $\Delta T$ is the greatest changes in temperature; $\Delta L$ is the length of tissue, and $A$ is the tissue level and $\Delta t$ exposure time. In the case of small lengths, the equation wi
ll be as follows:
\begin{equation}
\label{eq3}
f=-k\Delta T
\end{equation}
Heating conductivity inside tissue is obtained through biological heating equation \ref{eq4}. In this equation, $p$ is tissue density, $c$ is specific heat coefficient, $q_{s}$ is absorption of HIFU energy, qp is the rate of heat flow through the blood, and $q_{m}$ is indicative of metabolic activity.
\begin{equation}
\label{eq4}
 \rho c  \frac{\partial T}{\partial t} =  \bigtriangledown (k \bigtriangledown T)+ q_{s} + q_{p} + q_{m} 
\end{equation}
The best way to investigate heating effects is to calculate the ultrasound absorption rate, blood flow and metabolic activity. Factors such as blood density, specific heat, and temperature of blood affect blood flow.
In Figure \ref{fig3}, a general view of the heating of a capillary network and its parameters is displayed. The heating energy of a capillary network using equation 5 is obtained.
\begin{equation}
\label{eq5}
q_{p} = -w_{b} \rho_{b} c_{b}  \delta_{t}(T_{f}-T_{i})
\end{equation}
\\
In this regard, $w_{b}$ is blood flow, $\rho_{b}$ is blood density, $c_{b}$ is specific heat of blood, $\delta_{t}$ is specific heat of tissue, $T_{f}$ is final heat, and $T_{i}$ is initial heat.
By placement of the relationship in equation \ref{eq4}, we will arrive at:\\
\begin{equation}
\label{eq6}
\rho c. \frac{\partial T}{\partial t}= \bigtriangledown (k \bigtriangledown T) + q_{s} - w_{b} \rho_{b} c_{b}  \delta_{t}(T_{f}-T_{i})+q_{m}
\end{equation}
In the above equation, metabolic activity is negligible in front of the heat generated, and we will arrive at:
\\
\begin{equation}
\label{eq7}
\rho c. \frac{\partial T}{\partial t}= \bigtriangledown (k \bigtriangledown T) + q_{s} - w_{b} \rho_{b} c_{b}  \delta_{t}(T_{f}-T_{i})
\end{equation}
\\
According to Equation \ref{eq7}, the amount HIFU absorbed energy due to the model designed for tissue can be modified. The energy is associated with the similarity of tissue, density, and other biological factors. Usually when treatment is done, treatment points should be shown separately in separate images, and the operator should be specify separately that the same tissue is shown in Figure \ref{fig4}. Bright points represent treatment carried out by ultrasound.
In the capsule area, the temperature in the central and parietal regions is different than that shown in Figure \ref{fig5} \cite{habash2007thermal}.
As previously noted, the removal of unhealthy tissue by two phenomena of increasing the temperature and the cavitation in tissues is done. In the meantime, in some cases, thermal images are taken that have special importance. Below, thermal images before and after the treatment are shown. At the time of treatment, usually images will be thermographic (Figure \ref{fig6}) so that we can note the difference before and after treatment.

\subsection{Software Modeling}
\begin{figure}[h!]
	\includegraphics[width=\linewidth]{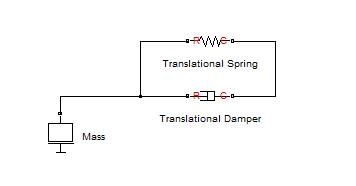}
	\caption{Selected model in simulation}
	\label{fig7}
\end{figure}

In this model, each layer of tissue using three elements of spring, damper and mass, as shown in Figure \ref{fig7}, is shown. In order to define the direction of modeling used to measure elasticity, viscosity and mass of tissue of elements, spring and damper are placed in parallel with each other, and mass is shown in a series with these two in circuit \cite{hajian2015new}.

\begin{figure}[h!]
 	\includegraphics[width=\linewidth]{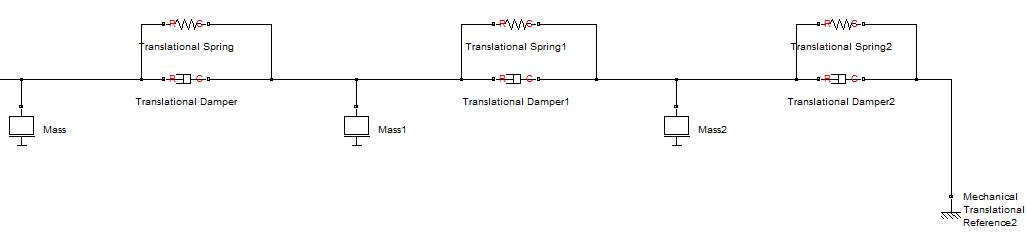}
 	\caption{Model for three layers}
 	\label{fig8}
 \end{figure}

 In this model, a three-layer tissue is intended. Therefore, these three elements, as shown in Figure \ref{fig8}, are repeated, and the last layer is connected to the ground.
 \subsubsection{Input Signal}
At the beginning, an ultrasound wave enters the tissue, where we define an input signal which consists of three pulses with a maximum range of 10 volts (Figure 9).
 \begin{figure}[h!]
 	\includegraphics[width=\linewidth]{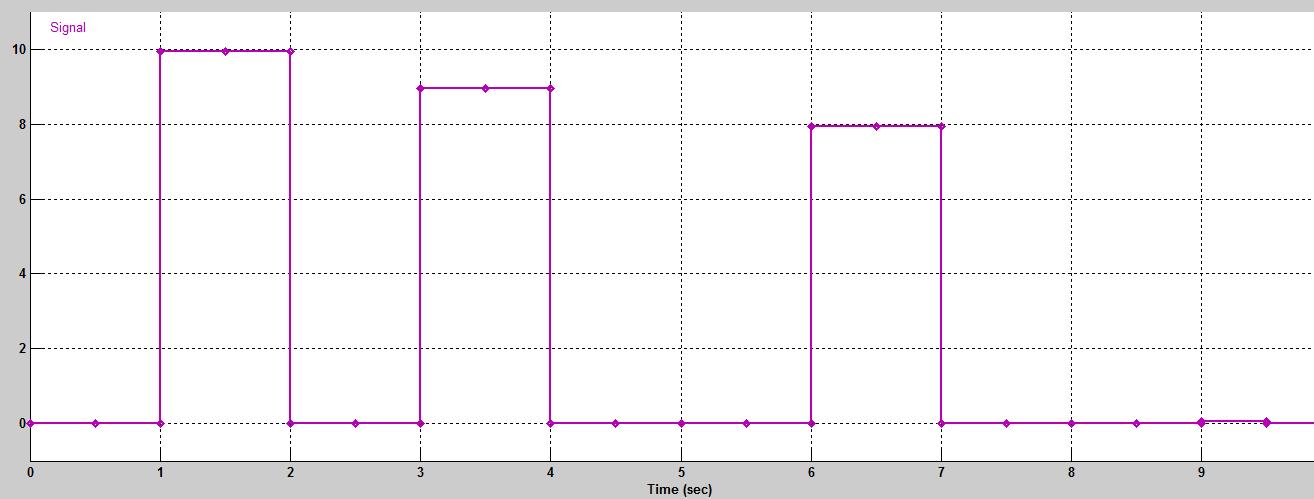}
 	\caption{Square wave designed to simulate}
 	\label{fig9}
 \end{figure}

\subsection{Simulation results}
In this paper, a linear method is used for simulation, and the mass, spring and damper model is expressed. The HIFU simulator is used to measure HIFU, a phantom, as Figure 10 is considered. By putting the values of the mass, spring and damper in the simulation, pressure of HIFU on a tissue can be achieved, including a layer of the kidney.
\begin{figure}[h!]
	\includegraphics[width=\linewidth]{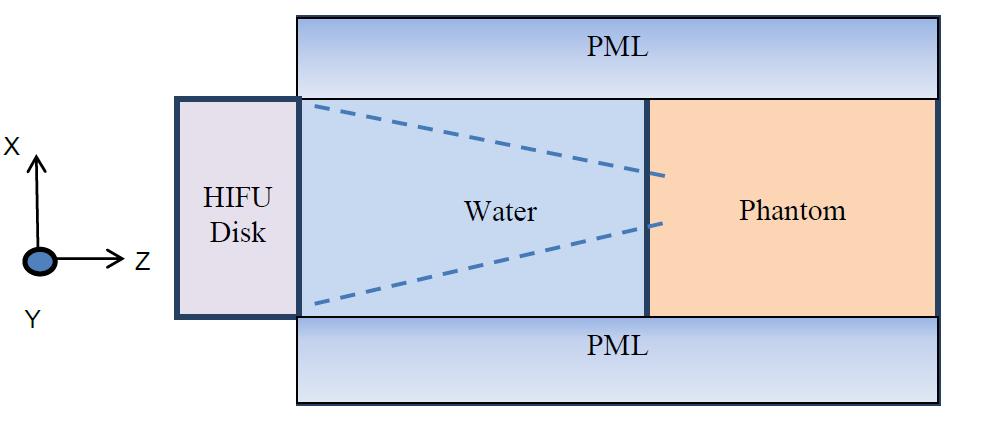}
	\caption{Figure of phantom intended for related software.}
	\label{fig10}
\end{figure}
Then, using liver values, the appropriate software estimates for heat and pressure are applied in different layers.
In the case of radiation from three directions and heat in the target point to liver tissue on the tumor (target), a temperature as high as 81 C for the modeling will be obtained. 
In this case, paths of radiations are considered, as shown in Figure \ref{fig11}. Liver and kidney comparative graphs of temperature as a result of HIFU radiation in the linear part are shown in three layers as Table \ref{table1}. Between the practical and simulator test, little difference was observed.
\begin{figure}[h!]
	\includegraphics[width=\linewidth]{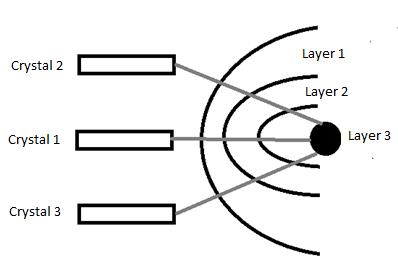}
	\caption{Test and design done to extract heat and pressure to liver tissue}
	\label{fig11}
\end{figure}
\begin{table}[htb]
	
	\centering
	\caption{Practical temperatures achieved for the first path and liver and kidney tissue viscoelasticity}
	\label{table1}
	\resizebox{\columnwidth}{!}{%
	\begin{tabular}{|c|c|c|c|}
		\hline
		\multicolumn{1}{|l|}{}          & \begin{tabular}[c]{@{}c@{}}Temperature of layer 1\\ (C)\end{tabular} & \begin{tabular}[c]{@{}c@{}}Temperature of layer 2\\ (C)\end{tabular} & \begin{tabular}[c]{@{}c@{}}Temperature of layer 3\\ (C)\end{tabular} \\ \hline
		Crystal Path 1 (liver)          & 47                                                                   & 39                                                                   & 27                                                                   \\ \hline
		Crystal Path 1 (kidney tissues) & 45                                                                   & 36                                                                   & 30                                                                   \\ \hline
	\end{tabular}
}
\end{table}
\\
In Figure \ref{fig11}, through a 50 percent reduction in the amount of viscoelasticity using the same path, the temperature of the first through third layer is obtained (Table 2).
\begin{table}[htb]
	\centering
	\caption{Temperature obtained in the second crystal path for the same path and 50\% reduction in liver tissue viscoelasticity}
	\label{table2}
	\resizebox{\columnwidth}{!}{%
	\begin{tabular}{|l|c|c|c|}
		\hline
		& \begin{tabular}[c]{@{}c@{}}Temperature of layer 1\\ (C)\end{tabular} & \begin{tabular}[c]{@{}c@{}}Temperature of layer 2\\ (C)\end{tabular} & \begin{tabular}[c]{@{}c@{}}Temperature of layer 3\\ (C)\end{tabular} \\ \hline
		\multicolumn{1}{|c|}{Crystal Path} & 41                                                                   & 30                                                                   & 22                                                                   \\ \hline
	\end{tabular}
}
\end{table}
\\
In this case, in the target point a 76 $^{\circ}$ C temperature, and for the second path a temperature of the first layer of 41 $^{\circ}$ C, and for the second and third paths 30 and 22 $^{\circ}$ C, respectively, are achieved. This issue proves the effects of viscoelasticity on the temperature and the pressure, and the relative change in the target temperature is 6 $^{\circ}$ C and the first layer is 14\%. Performing this issue for pressure causes it to create relative changes of 9 percent as a result of changing viscoelasticity of a path. Measurement in this simulation is done by removing a living tissue from the body. If it is in any path, the value of the physical parameter is measured, and biophysical tissue damage in the same path is prevented as a practical matter. For this purpose, the volume of contour between the HIFU probe to the target in the included K layer and each layer is composed of a voxel matrix, and viscoelastic elements should be extracted in a non-invasive manner (imaging) and placed in the designed model, which is a different method. Additionally, if we want the temperature at the target point to remain at a certain level, mechanical properties can be changed. Heat at a given point or the excitation signal of the HIFU probe is modified, and by this action, inverse treatment planning will occcur.
\begin{figure}[h!]
	\includegraphics[width=\linewidth]{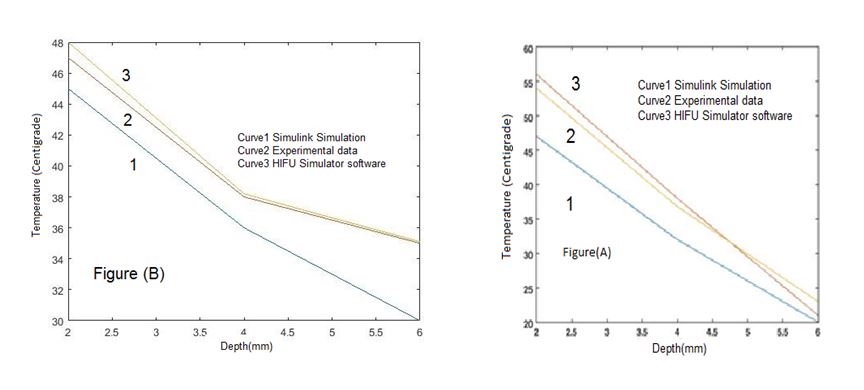}
	\caption{Compare the temperatures of the liver (Figure A) and kidney (Figure B) due to wave propagation of the HIFU in depth in various modes}
	\label{fig12}
\end{figure}
As shown in Figure \ref{fig12}, the percent of relative change of temperature measured in the first, second and third layers for the designed model to practical mode was obtained, and it was at 4, 5, and 14. Using the simulator HIFU, 6, 5.7, and 14.5 percent were obtained. The values for the liver tissue for the practical mode in the first and third layers, respectively, is 16, 17, and 24\%, and relative changes using the simulator HIFU are 12.9, 17.9, and 15 percent.
A comparison chart of liver and kidney tissue pressure, described in Figure 13, is displayed. The results of this figure in two modes using the software of simulator HIFU and model simulation have been carried out, and the following numbers were achieved: 14.59 and 13.39 MPa. In calculating for pressure mode, 9\% relative changes are observed. It should be noted that in pressure mode between the liver and kidneys, no significant changes were observed.
\begin{figure}[h!]
	\includegraphics[width=\linewidth]{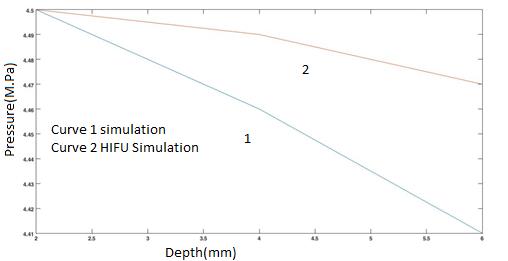}
	\caption{Compare the liver and kidney pressures due to wave propagation of HIFU in depth in various positions}
	\label{fig13}
\end{figure}

\subsection{Conclusion}
HIFU is a method that is used primarily for the removal of solid tumors through the creation of heat. This method requires the deliberate design of appropriate treatment. In this way, designing the appropriate treatment prevents damage to healthy tissues and results in the best treatment for the patient. Hence, it is very important. In this article, we sought to investigate pressure changes in kidney and liver tissue by modeling and stimulating HIFU processes to determine optimal values, to prevent complications, and to provide the optimal treatment method. The model and assumption represent a new method in treatment. Planning and precision in this case are also significant factors when comparing the HIFU simulator. The advantages of this model can be attributed to proper temperature distribution based on medical need and lack of skin burns caused by sound mechanical effects. Additionally using this method causes set temperature to be employed at specific limits, and as a result, normal tissue is not damaged.
\\
The presence of blood effects and pulse can be considered as pulse; in this case, the mechanical elements will be changed with time. In addition, blood changes the heat distribution in the tissue, to overcome this problem, a correction factor should be considered. In this article, the correction factor of stimulation as an electrical wave in the scale of MV was considered to the input of tissue that in Simulink simulation was not isolated from ultrasound stimulation.
\\
\\

\end{document}